\documentclass[prl,letterpaper,twocolumn,showpacs,superscriptaddress]{revtex4}
\usepackage{graphicx,psfrag,amsmath,amssymb,amsfonts,bbm,latexsym,color,dcolumn}

\begin{document}

\title{Are quantum measurements on macroscopic bodies feasible?}

\author{Roberto Onofrio}

\affiliation{Scuola del Dottorato di Ricerca, Dipartimento di Fisica
  dell'Universit\`a di Roma ``La Sapienza'', Roma ,Italy}

\date{15 April 1990, published in {\sl Europhys. Lett.} {\bf 11} (1990) 695}

\begin{abstract}
The possibility to test quantum measurement theories is discussed in
the more phenomenological framework of the quantum nondemolition
theory. A simple test of the hypothesis of the state vector collapse 
is proposed by looking for deviations from the Boltzmann distribution 
of the energy associated to only one complex amplitude of a
macroscopic oscillator.

\end{abstract}


\maketitle

The long debate on the foundations of quantum mechanics and the
Copenhagen interpretation gave rise to speculations on the possibility
of statistical interpretations of this theory. In particular the
necessity for a postulate on the collapse of the state vector after a
measurement has also been discussed. Alternative quantum measurement
theories which are able to recover all the statistical results of the 
orthodox approach without such a postulate have been developed
\cite{EPL1}. The success of standard quantum mechanics and the
impossibility to experimentally distinguish it from more refined 
approaches without individual interpretation or absence of postulates
on the effect of the measurement process on the state vector is
related to the field of application of such a theory, namely the
microscopic world. All the empirical facts known on the microscopic
world are based on averaged measurements on ensemble of particles 
and in this way the two approaches, statistical (Einstein,
Schroedinger) as well as individual (Born, Bohr), are ``degenerate'',
{\it i.e.} they cannot be distinguished. A possible discrimination
between these two interpretations of quantum mechanics may be obtained
by measuring in a quantum regime of sensitivity some observable of a
single macroscopic body such as, for instance, the position. This
possibility has not been seriously taken into account for a long time,
due to the experimental difficulties of high-precision measurements of
position. At the same time this drawback has been the basis of the
understanding of the classical limit, as usually shown in the standard
textbooks of quantum mechanics. Indeed, in a typical macroscopic body,
mainly due to the relevant mass, the action is very large compared to
the natural unit for it in quantum mechanics, the Planck constant. 
Thus, provided that the classical sources of noise are large, a
classical schematization for all the macrophysics systems, and in
particular for the measuring apparatus, is enough.

On the other hand, there has been considerable progress in the
measurement technology of small values of the action through the
reduction of the classical sources of noise, mainly motivated by the
stringent requirements for high-precision measurements of space-time
in experimental gravitation to test post-Newtonian effects
\cite{EPL2,EPL3}. All this has given rise to a new field of research
consisting in the study of the effect of a quantum-limited read-out on
the state of a macroscopic harmonic oscillator \cite{EPL4,EPL5,EPL6,EPL7}.

In this letter I would like to discuss quantum-limited measurements on
macroscopic oscillators concerning in particular the possibility to
obtain results of interest on quantum theory of measurement with a
dedicated experimental set-up and a proper strategy of measurement. 
The limitations which quantum mechanics gives on the sensitivity of 
a position measurement have been analyzed in detail from various 
points of view in the framework of a quantum-mechanical formalism 
including the postulate on the state vector reduction \cite{EPL8}. 
The existence, in such a framework, of the so-called quantum 
nondemolition (QND) observables has been recognized \cite{EPL5,EPL6}. 
By implementing a QND strategy, one can repeatedly monitor an 
observable of a quantum system without affecting the predictions 
of the new result on the next measurements. In order to do
this, some conditions, on both the observable and the measurement
strategy, have to be satisfied. We recall (see \cite{EPL6} for an
overview) that a QND measurement of the observable $\hat{X}_\alpha$ is
obtained provided that the interaction Hamiltonian $\hat{H}_i$ between
the measured and the measuring systems satisfies 
\begin{equation}
[\hat{H}_i, \hat{X}_\alpha]=0.
\end{equation}
Furthermore, only particular observables can be repeatedly monitored
without affecting the output of a new measurement, and the required
condition is that the observable commutes with itself at the different
instants during which measurement are done
\begin{equation}
[\hat{X}_\alpha(t_i),\hat{X}_\alpha(t_j)]=0.
\end{equation}
Whenever (1) ad (2) are satisfied QND strategies of measurement may be
performed. The physical meaning of QND measurements on the position of
a macroscopic body is then obtained by thinking of the following
procedure: in a first measurement the state is prepared, according to
the state vector postulate, in the eigenstate of the obtained
eigenvalue. The next measurements are all exactly predictable, due to
(2), on the basis of the Heisenberg evolution, provided that the
interaction Hamiltonian does not affect he monitored observable
according to (1). A possible discrepancy between the predicted
measurements and the ones really done gives quantitative indications
on the change of the state as that due to an external force, {\it
  i.e.} the original goal of the QND techniques. Furthermore, despite
the quantum regime and due to their particular nature, QND strategies
give rise to well-defined measurements also on a single system. This
it is possible to overcome the requirements on a statistical ensemble
of measured objects usually done for quantum systems; only one system
is enough to obtain deterministic predictions and measurements. In
this sense observable monitored with a QND strategy have a behaviour
completely similar to their classical counterparts or, in other words,
they constitute a classical subset between all the possible
measurements in a quantum regime. This point, although it seems
paradoxical, has already been stressed in other papers (for example in
\cite{EPL9}).

These techniques are currently implemented, although in a fully
classical regime, on real experiments by various gravitational groups
in view of their use on the third generation of gravitational wave
antennas \cite{EPL10,EPL11,EPL12,EPL13,EPL14}. Two questions related
to this last remark arise: how far are we from a quantum-limited noise
scenario for a real experiment? How is it possible to test the wave
function collapse hypothesis by means of such techniques?

The answer to the first point requires the introduction of some
notions on the detection of small diplacements of macroscopic
oscillators, although we refer to \cite{EPL7} for a detailed
description of the so-called electromechanical transducers. A general
scheme for the detection of small forces consists at least of a
mechanical oscillators, an electrical oscillator, an electromagnetic
reservoir (also called pump) which allows correlations between the
time evolution of the two oscillators (it is responsible for the
interaction Hamiltonian), and an amplifier. All these elements are
characterized by well-known classical sources of noise. Brownian and
Johnson noise in the oscillators, amplitude and phase noise in the
electromagnetic reservoir (pump noise) and amplifier noise. The
Brownian noise is reduced at a low level by cooling the mechanical
oscillator at now temperature. The average energy variation due to the
Brownian drift in a time $\Delta t$, referred to the quantum of energy
$\hbar \omega_1$ of a mechanical oscillator having frequency
$\omega_1/2\pi$, is equal to 
\begin{equation}
\eta_1=\frac{\Delta E_\mathrm{br}}{\hbar
  \omega_1}=\frac{k_\mathrm{B}T}{\hbar \omega_1} \frac{\Delta t}{\tau_1}
\end{equation}
\noindent
$T$ being the thermodynamical temperature and $\tau_1$ the relaxation
time of the mechanical oscillator. The next generation of
gravitational-wave antennas is planned to operate at a temperature of
$(50 \div 100)$ mK by means of dilution refrigerator facilities
\cite{EPL15}. In this situation, typical values available for the
previous parameters are $T \simeq (50 \div 100)$ mK, $\omega_1 \simeq
10^4 $ s$^{-1}$, $\Delta t \simeq 10^{-2}$ s, $\tau \simeq (10^3 \div
10^4)$ s giving $\eta_1 \simeq 1$. Thus the transition probability
between different energy levels due to quantum uncertainty begins to
be comparable to the transition probability due to thermal
fluctuations. The same analysis may be repeated for a low dissipation
(superconducting) electric circuit obtaining a similar noise figure
\begin{equation}
\eta_2=\frac{k_\mathrm{B}T}{\hbar \omega_2} \frac{\Delta t}{\tau_2}
\end{equation}
\noindent
with $\eta_2 \simeq 1$ by assuming $T \simeq (50 \div 100)$ mK, $\omega_2 \simeq
10^7 \div 10^8$ s$^{-1}$, $\Delta t \simeq 10^{-2}$ s, $\tau \simeq 1$
s. The amplifier noise may be very low, of the order of few noise
quanta $\eta_a= 1\div 10$, by using SQUIDs or GaAs amplifiers
\cite{EPL16,EPL17,EPL18}. The current troubles come from the amplitude
and phase noise and in order to overcome this limitation balanced
bridge configurations for the electrical circuit are used
\cite{EPL10,EPL19}. However the corrections to the maximum squeezing
due to the pump quantum fluctuations have been already calculated by
using the second quantization formalism \cite{EPL20}. Therefore it
would be clear that a refined technology is available which, although
originally developed in the context of the gravitational wave
detection, is mature enough to perform real measurements on
macroscopic oscillators near the quantum limit. Furthermore in an
experiment dedicated to test QND measurements some typical constraints
imposed on the design of a gravitational wave detector may be
relaxed. For instance it is possible to choose higher mechanical
frequencies, improving the acoustic insulation of the apparatus, small
masses in order to increase the zero point displacements ($\Delta x
\simeq {(\hbar/m\omega)}^{1/2}$), very low temperatures by using
adiabatic demagnetization, already available down to ($10^{-4} \div
10^{-5}$) K for small times and small masses \cite{EPL21}. Small
values for the test masses are also welcome for the application of 
atomic force microscopy which begins to be used for high-sensitivity
measurements of positions \cite{EPL22,EPL22,EPL24}.

Now we analyze the second question, {\it i.e.} how to obtain
information on the possible existence of the wavefunction collapse. As
appears from the previous discussion on the sources of noise, the
macroscopic oscillator is in thermal equilibrium with an external
reservoir. The effect of the reservoir can be taken into account by
simply adding to the Hamiltonian of the system a term which expresses
the effect of the Brownian stochastic force. This stochastic force is
responsible for the coupling of the system to the external world and
gives rise to a well-defined energy distribution in which a
temperature may be introduced. Indeed it is well known that the
classical analysis of a force acting on an oscillator at thermal
equilibrium is done by looking for the energy distribution $\rho(E)$
(see for instance \cite{EPL7}). This distribution, at thermal
equilibrium and in the absence of external forces of nonstochastic
nature, is given by the Boltzmann relation 
\begin{eqnarray}
\rho(E)& = &\alpha \exp{(-E/k_B T)} \nonumber \\
& = & \alpha \exp{[-m\omega_1^2(X_1^2+X_2^2)/2k_BT]},
\end{eqnarray}
\noindent
where, in the last expression, the distribution has been written in
terms of the two complex amplitudes of the harmonic oscillator $X_1$
and $X_2$ defined as
\begin{eqnarray}
X_1 & = & Re\left[\left(x+i \frac{p}{m\omega_1}\right) \exp{[i
      \omega_1t]}\right], \nonumber \\
X_2 & = & Im\left[\left(x+i \frac{p}{m\omega_1}\right) \exp{[i \omega_1t]}\right].
\end{eqnarray}
\noindent
It is possible to prove that the complex amplitudes of a harmonic
oscillator are QND observables \cite{EPL6}. A clear prediction of the
QND theory is that, whatever the value of the thermodynamical
temperature, the energy distribution for the $X_1$ degree of freedom
monitored in a QND way remains Boltzmann-like also in a quantum
regime, {\it i.e.}
\begin{eqnarray}
\rho_1(E_1) & = & \alpha \cdot \exp{(-E_1/k_B T_1)} \nonumber \\ 
& = & \alpha \cdot \exp{(-m \omega_1^2 X_1^2/2k_B T_1)},
\end{eqnarray}
\noindent
where $T_1$ is the effective temperature for the phase $X_1$ only due
to the Brownian motion. This is of course due to the absence of
effects on $X_1$ of the quantum noise induced in the measurement, this
last affecting only the complex amplitude $X_2$. This means that,
despite the quantum regime, the statistics for the degree of freedom
$X_1$ measured in a QND way is classical. A more detailed discussion
of the classical properties of the QND measurements is given in
\cite{EPL9} where the authors apply the Feynman path integral
formalism in order to have the classical limit always in evidence. 
The energy distribution for $X_2$ will be unpredictable on the pure
basis of statistical mechanics, being affected by the quanta randomly
introduced during the measurement process. Thus, no effective
temperature for the quantum-demolished phase $X_2$ may be introduced,
due to the highly nonequilibrium states induced by the quantum during
the measurements. The observation of deviations from the Boltzmann
distribution (7) for $X_1$ below a given temperature, provided that it
does not arise from other sources as for example external forces, will
give indications on possible departures from the collapse of the state
predicted by the orthodox approach on which QND theory is based. In
other words, if the Copenhagen interpretation is true QND observables
exist and it will therefore be possible to observe a classical
distribution of the energy associated to the measurement of these
observables. 

It is thus evident that there already exists a phenomenological
scenario in which quantum measurement theorists may quantitatively
apply their ideas by computing predictions for a real experiment. On
the other hand, it is also clear that, in order to achieve such a
goal, all the noise figures previously cited have to be obtained
simultaneously within the same set-up and this is surely a difficult
task. 

However this scenario, if properly developed, can give good prospects
for small-scale dedicated experiments on QND measurement
strategies. At the same time this framework gives rise to interesting
purely theoretical inquiries about the determination of the energy
distribution of a macroscopic oscillator on which measurements of
general kind, not necessarily of QND nature, at a quantum sensitivity
are performed having, at the same time, a weak thermal equilibrium due
to fluctuation-dissipation forces. Furthermore it is interesting to
analyse, both theoretically and experimentally, the time reversal in
such a kind of systems. In this way it would be possible to understand
if the two kinds of irreversibility today encountered in physics, {\it
  i.e.} the one coming from statistical mechanics and the other
originating from the measurement process in quantum mechanics, have
the same origin according to a conjecture of Landau and Lifshitz
\cite{EPL25} and as it is further formally suggested by the Feynman
approach to quantum mechanics and quantum field theory \cite{EPL26}. 

\acknowledgments
I wish to acknowledge M. F. Bocko, F. Bordoni, and S. Nicolis for
useful discussions. This work was supported by the INFN, Sezione di
Roma, Italy.


\begin{thebibliography}{99}

\bibitem{EPL1} For a general review on the subject see for example
  Wheeler J.A. and Zurek W.H. (Editors), {\sl Quantum Theory and
  Measurement} (Princeton University Press, New York, N.Y.) 1983 and
  the reference cited therein.

\bibitem{EPL2} Braginskii V.B., Caves C.M., and Thorne K.S.,
  {\sl Phys. Rev. D}, {\bf 15} (1977) 2047.

\bibitem{EPL3} Thorne K.S., {\sl Rev. Mod. Phys.}, {\bf 52} (1980) 285.

\bibitem{EPL4} Braginskii V.B., {\sl Sov. Phys.JETP}, {\bf 26} (1968) 831.

\bibitem{EPL5} Braginskii V.B. and Vorontsov Yu., {\sl Sov. Phys. JETP}, {\bf 17} (1975) 644.

\bibitem{EPL6} Caves C.M., Thorne K.S., Drever R.P., Sandberg V.D. and
  Zimmermann M., {\sl Rev. Mod. Phys.}, {\bf 52} (1980) 341.

\bibitem{EPL7} Braginskii V.B., Mitrofanov V.P. and Panov V.I.,
  {\sl Sistemi s maloi dissipatsiei} (Nauka, Moscow) 1981 
[{\sl Systems with Small Dissipations} (University of Chicago,
  Chicago, Ill.) 1985].

\bibitem{EPL8} Bohm D., {\sl Quantum Theory} (Prentice-Hall, Englewood
  Cliffs, N.J.) 1951, chapts. 6 and 22. 

\bibitem{EPL9} Golubsova G.A. and Mensky M.B., {\sl Int. J. Mod. Phys. A}, {\bf 4} (1989) 2733. 

\bibitem{EPL10} Bocko M.F. and Johnson W.W., {\l Phys. Rev. Lett.},
  {\bf 47} (1981) 1184; {\bf 48} 1982) 1371.

\bibitem{EPL11} Blair D., {\sl Phys. Lett. A}, {\bf 91} (1982) 192.  
A. V. Plyukhin, Europhys. Lett. \textbf{75}, 15 (2006).

\bibitem{EPL12} Spetz G.W., Mann A.G., Hamilton W.O. and Oelfke W.C.,
  {\sl Phys. Lett. A}, {\bf 104} (1984) 2135.

\bibitem{EPL13} Bordoni F., De Panfilis S., Fuligni F., Iafolla V. and
  Nozzoli S., in {\sl Proceedings of the IV Marcell Grossmann Meeting
  on General Relativity}, edited by R. Ruffini (Elsevier, Amsterdam)
  1986, p. 353.

\bibitem{EPL14} Barro E., Onofrio R., Rapagnani P. and Ricci F., in 
{\sl Proceedings of the I International Symposium on Experimental
  Gravitational Physics}, edited by P.F. Michelson (World Scientific,
Singapore) 1988, p. 363.

\bibitem{EPL15} Amaldi E. {\it et al.}, in {\sl Proceedings of the V
  Marcell Grossmann Meeting on General Relativity}, to be published.

\bibitem{EPL16} Koch R.H., Von Harlingen D.J. and Clarke J., {\sl
  Appl. Phys. Lett.}, {\bf 38} (1981) 380.

\bibitem{EPL17} Weinreb S., {\sl IEEE Trans. on Microwave Theory and
  Techniques}, MTT-28, no. 10 (1980) 1041.

\bibitem{EPL18} Bocko M.F., {\sl Rev. Sci. Instrum.}, {\bf 55} (1984)
  256; Bordoni F., private communication. 
 
\bibitem{EPL19} Fuligni F and Iafolla V., in {\sl Proceedings of the
  III Marcell Grossman Meeting on General Relativity}, edited by Hu
  Ning (Elsevier, Amsterdam) 1983, p. 1489.

\bibitem{EPL20} Crouch D.D. and Braunstein S.L., {\sl Phys. Rev. A},
  {\bf 38} (1988) 4696. 

\bibitem{EPL21} Richardson R.C. and Smith E.N. (Editors), {\sl
  Experimental Techniques in Condensed Matter Physics at Low
  Temperatures} (Addison-Wesley, New York, N.Y.) 1988.

\bibitem{EPL22} Niksch M. and Binnig G., {\sl J. Vac. Technol. A},
  {\bf 6} (1988) 470.

\bibitem{EPL23} Bocko M.F., Koch R.H. and Stephenson K.A., {\sl
  Phys. Rev. Lett.}, {\bf 61} (1988) 726. 

\bibitem{EPL24} Bordoni F., Fuligni F., Bocko M.F. and Koch R.H., in 
{\sl Proceedings of the International Workshop on GW Analysis and
  Processing, Amalfi}, July 1-5, 1988, to be published.

\bibitem{EPL25} Landau L.D. and Lifshitz E.M., {\sl Statistical
  Physics} (Addison-Wesley, Reading, Mass.) 1969.

\bibitem{EPL26} Feynman R.P.and Hibbs A.R., {\sl Quantum Mechanics and
  Path Integrals} (McGraw-Hill, New York, N.Y.) 1965.

\end{thebibliography}
\end{document}